\documentclass [prc,aps,showpacs]{revtex4}
\usepackage{graphicx}
\begin{document}

\newcommand{\adag}{a^{\dag}}
\newcommand{\atil}{\tilde{a}}
\def\frp*1{${*1\over2}^+$}
\def\frm*1{${*1\over2}^-$}
\def\g{\noindent}
\def\mev{\hbox{\MeV}}
\def\kev{\hbox{\keV}}
\def\lambdabar{{\mathchar'26\mkern-9mu\lambda}}
\def\lambdabarrr{{^-\mkern-12mu\lambda} }

\title{Microcanonical Ensemble Extensive Thermodynamics of Tsallis Statistics}
\author{A.S.~Parvan}


\affiliation{Bogoliubov Laboratory of Theoretical Physics, Joint
Institute for Nuclear Research, 141980 Dubna, Russia}
\affiliation{Institute of Applied Physics, Moldova Academy of
Sciences, MD-2028 Kishineu, Republic of Moldova}

\begin{abstract}
The microscopic foundation of the generalized equilibrium
statistical mechanics based on the Tsallis entropy is given by
using the Gibbs idea of statistical ensembles of the classical and
quantum mechanics. The equilibrium distribution functions are
derived by the thermodynamic method based upon the use of the
fundamental equation of thermodynamics and the statistical
definition of the functions of the state of the system. It is
shown that if the entropic index $\xi=1/(q-1)$ in the
microcanonical ensemble is an extensive variable of the state of
the system, then in the thermodynamic limit
$\tilde{z}=1/(q-1)N=\mathrm{const}$ the principle of additivity
and the zero law of thermodynamics are satisfied. In particular,
the Tsallis entropy of the system is extensive and the temperature
is intensive. Thus, the Tsallis statistics completely satisfies
all the postulates of the equilibrium thermodynamics. Moreover,
evaluation of the thermodynamic identities in the microcanonical
ensemble is provided by the Euler theorem. The principle of
additivity and the Euler theorem are explicitly proved by using
the illustration of the classical microcanonical ideal gas in the
thermodynamic limit.
\end{abstract}



\maketitle

\section{Introduction}
The equilibrium statistical mechanics and thermodynamics are well
defined theories in modern physics~\cite{Gibbs,Balescu}.
Applications of these theories are restricted by investigation of
the so-called thermodynamic or statistical systems which are
constrained by several rigid requirements~\cite{Kvasn}. One of the
first attempts to construct the generalized equilibrium
statistical mechanics based on the mathematical redefinition of
the Boltzmann-Gibbs statistical entropy and the principles of the
information theory belongs to C.~Tsallis~\cite{Tsal88}. Until
recently, there has been a great deal of interest in studying
nonextensive thermodynamics due to its relevance in many fields of
physics~\cite{Tsal99,Gudima}. However, many fundamental features
regarding the violation of the zero law of thermodynamics and the
principle of additivity remain unclear~\cite{Abe0,Parv1}. Note
that these difficulties have resulted in the occurrence of a large
number of variants of the Tsallis generalized statistical
mechanics~\cite{Tsal98}.

The statistical mechanics investigates thermodynamic systems which
are defined solely by the specification of macroscopic variables
on the basis of the theory of probability and the microscopic laws
of the classical and quantum mechanics. The evolution of the
macroscopic system with a large number of degrees of freedom is
impossible to describe by only dynamic methods. Therefore, the
Gibbs idea of statistical ensembles is usually used~\cite{Zub}.
All information about the macrostate of the system is contained in
the phase distribution function, which evolves according to the
Liouville equation or in the statistical operator whose evolution
with time is described by the von Neumann equation. To derive the
phase distribution function and the statistical operator is the
primary goal of the nonequilibrium statistical mechanics.

In particular, the equilibrium statistical mechanics implies that
one uses the Gibbs equilibrium statistical ensembles. In the state
of thermodynamic equilibrium of the system, the phase distribution
function and the statistical operator do not depend on time.
Therefore, they are functions only of the first integrals of
motion of the dynamic system. In this case, the mechanical laws
and the Liouville and von Neumann equations do not allow one to
determine unequivocally the equilibrium distribution function and
the statistical operator~\cite{Balescu,Zub}. Therefore, an obvious
dependence of the equilibrium distributions on the macroscopic
variables of the state of the system is defined by introducing
additional postulates. The traditional way is based on the Gibbs
postulate of the equiprobability of the dynamic states of the
isolated system~\cite{Gibbs}. The alternative way rests on the
Jaynes principle of a maximum of the information
entropy~\cite{Jaynes}. The statistical mechanics constructed on
the Gibbs equilibrium distributions, which corresponds to the
Boltzmann-Gibbs statistical entropy, completely satisfies all
postulates of the equilibrium thermodynamics.

Standard treatments of the Tsallis statistics point out that the
entropic index $q$ is an additional intensive parameter, which has
a fixed value for different thermodynamic systems~\cite{Tsal98}.
This concept leads to shortcomings of thermodynamics and needs to
be reconsidered. As shown further, these problems can be resolved
by the assumption that the parameter $\xi=1/(q-1)$ is the
extensive argument of the statistical entropy.

The paper is organized as follows. In the second section, the
microscopic foundation of the Tsallis generalized statistical
mechanics is given. The microcanonical equilibrium distribution
function and statistical operator are deduced in the third
section. In the fourth section, the performance of the
thermodynamic principles in the microcanonical ensemble of the
Tsallis statistics are proved. The developed formalism is
exemplified in the fifth section by treating the classical
microcanonical ideal gas.

\section{Microscopic Foundation of Tsallis Statistical Mechanics}
A macrostate of a system with a large number of degrees of freedom
is imperfectly known at the microscopic level. The system can be
found in any dynamic state compatible with the external
macroscopic conditions. Therefore, for the macroscopic system it
is possible to maintain only the probabilistic description of
dynamic processes. For this reason, in the statistical mechanics
the Gibbs idea of statistical ensembles is straightforward. The
macroscopic state of the system is represented as a set of a large
number of copies of the dynamic system under identical macroscopic
conditions. Each system of the ensemble is represented by a point
in phase space. Any physical observable $A$ of the macroscopic
system is represented as the expectation value $\langle A
\rangle^{t}$ of the dynamic variable $A(x,p,t)$ with the phase
distribution function $\varrho(x,p,t)$. The evolution with time of
a phase distribution function is governed by the Liouville
equation. According to the Liouville theorem, the volume of a
region in phase space remains constant in the process of movement
of phase points. The phase distribution function is constant along
the phase trajectories, $\varrho(x,p,t) = \varrho (x',p',t')$. To
describe the quantum many-particle systems the mixed states are
considered. A macrostate thus appears as a set of possible
microstates, which are set up by state vectors
$|\Psi_{r}(t)\rangle$, $r=1,2,\ldots$, each with its own
probability $w_{r}$ for its occurrence. The statistical operator
$\varrho(t)$ allows to determine the expectation value of a
dynamic variable $A$ regardless of the choice of the set of
quantum states $\{|\Psi_{r}(t)\rangle\}$. The evolution with time
of a statistical operator is governed by the von Neumann equation.
In the state of thermal equilibrium of the macroscopic system, the
phase distribution function $\varrho_{eq}(x,p)$ and the
statistical operator $\varrho_{eq}$ should not depend on the time
$t$. Therefore, from the Liouville and von Neumann equations it
follows that they are the first integrals of motion which must
depend only on the first constants of motion of the system.
Moreover, if these quantities are unequivocal and additive, then
there exist only four such integrals of motion: energy $H$, the
total momentum vector $\mathbf{P}$, the total angular momentum
vector $\mathbf{M}$, and the number of particles $N$.
In~\cite{Balescu,Zub}, the basic definition of the microscopic
foundation of the statistical mechanics is explained in detail.
The present investigation rests on this assumption.

The equilibrium distribution function and the equilibrium
statistical operator are not determined unequivocally by the
mechanical laws. To express the equilibrium distributions from the
macroscopic variables of state, the introduction of additional
postulates is required. A traditional way to construct the
equilibrium distributions is based on the Gibbs postulate of the
equiprobability of all accessible dynamic states of the isolated
system~\cite{Gibbs}. An alternative way for this is based on the
statistical definition of the entropy and the use of the Jaynes
principle explored in the information theory~\cite{Jaynes}. In the
present study, we suggest a new method based on the laws of the
equilibrium thermodynamics.

For this reason, we briefly recall the general laws of the
macroscopic equilibrium thermodynamics. The thermodynamic systems
are the object of the equilibrium thermodynamics and they must
satisfy some obligatory conditions~\cite{Kvasn}. First, these are
the systems of a large number of particles interacting with each
other and with external fields. Second, for every thermodynamic
system the zero law of thermodynamics is fulfilled, i.e., for such
a system there exists a state of thermal equilibrium, which
eventually is reached by the system at the fixed external
conditions. This principle guaranties the existence of the special
thermal measure, the temperature $T$, which is a general
characteristic of all thermodynamic systems in equilibrium contact
and which does not depend on the place and the method of
measurements.

Third, for thermodynamic systems the principle of additivity is
valid: all variables belong to two classes of additivity,
according to the reaction of a given physical one to the division
of the equilibrium system into the equilibrium macroscopic parts,
for example, into two parts. The extensive variables can be split
into two parts, and they should be proportional to the actual
amount of matter present, $\mathcal{F}_{1+2} =\mathcal{F}_{1} +
\mathcal{F}_{2}$. On the other hand, the intensive variables have
to keep its values and cannot depend on the size of the system,
$\phi_{1+2} = \phi_{1} = \phi_{2}$. As an example, we may consider
the thermodynamic systems which may be fixed in terms of the
macroscopic variables of state $T,V,N$. In this case, the
thermodynamic principle of additivity is implemented if intensive
quantities are functions of intensive arguments, and extensive
variables are proportional to the number of particles of the
system multiplied by the intensive quantity. Such dependence of
extensive and intensive variables is provided by the thermodynamic
limit~\cite{Kvasn}. In this respect, all expressions have to be
exposed to a formal limiting procedure
$N\rightarrow\infty,V\rightarrow \infty, v=V/N =\mathrm{const}$,
and only main asymptotics on $N$ should be kept. Then the
extensive variables $\mathcal{F}$ can be written $(\alpha> 0)$
\begin{equation} \label{106}
 \mathcal{F}(T,V,N)\left|_{N\rightarrow\infty \atop
      v = \mathrm{const}} \right. = N(f(T,v) + O(N^{-\alpha}))
      \stackrel{as} = N f(T,v),
\end{equation}
whereas the intensive variables $\phi$ take the following form:
\begin{equation} \label{107}
\phi(T,V,N)\left|_{N\rightarrow\infty \atop
      v = \mathrm{const}} \right. = \phi(T,v) + O(N^{-\alpha})
      \stackrel{as} = \phi(T,v),
\end{equation}
where $v = V/N$ is the specific volume and $f =\mathcal{F} /N$ is
the specific $\mathcal{F}$. Note that the thermodynamic limit is a
one-limiting procedure. The transitions not coordinated among
themselves $N\rightarrow\infty$ and $V\rightarrow\infty$ have no
physical sense, as in this case we would get results for either
the superdense system or the empty one.

Fourth, in relation to the thermodynamic systems the first, the
second and the third principles of thermodynamics are fulfilled,
being the mathematical basis of the macroscopic theory. The first
principle postulates the energy $E$ conservation law. The second
principle of thermodynamics in the axiomatic formulation of
R.J.~Clausius postulates the existence of a function of state
$S_{\mathrm{T}}$, called entropy. The absolute value of entropy is
determined from the third law of thermodynamics or the Nernst
theorem. The first and the second principles of thermodynamics for
the quasistatic reversible processes are combined to give the
fundamental equation of thermodynamics:
\begin{equation} \label{115}
  T dS_{\mathrm{T}} = dE + p dV + X dz - \mu dN,
\end{equation}
where $z =(z_{1},\ldots,z_{k})$ and $V$ are the "thermodynamic
coordinates"; $X = (X_{1},\ldots,X_{k})$ and $p$ play the role of
the associated "forces"; $\mu =\{\mu_{i} \}$ are the chemical
potentials and $N = \{N_{i} \}$ are the number of particles for
each kind $i$, respectively. The second law of thermodynamics for
nonequilibrium states, also formulated by R.J.~Clausius, refers to
the irreversible processes. This principle gives the direction of
a real process allowing one to investigate the properties of
equilibrium states as extreme ones. The most complete account of
the equilibrium and non-equilibrium processes and the role of the
characteristic times in the macroscopic thermodynamics is found
in~\cite{Kvasn,Prigogine}.

The expectation values of a dynamic variables with the equilibrium
distribution function and the equilibrium statistical operator
must satisfy all the postulates of the equilibrium thermodynamics.
The connection between the distribution function and the
macroscopic thermodynamical variables of the state is provided by
the statistical entropy. Usually it is determined on the base of
the information entropy. Let us define the Tsallis information
entropy, which recently has received wide popularity due to the
property of nonextensivity and which is used for construction of
the so-called generalized statistical
mechanics~\cite{Tsal88,Tsal98}. The Tsallis information entropy
for the discrete distribution of probabilities $\{p_{i} \}$ for
$W$ independent elementary events is defined in the following
manner~\cite{Tsal88}:
\begin{equation} \label{122}
S_{\mathrm{inf}} = -k \sum\limits_{i=1} ^{W} \frac{p_{i} -p_i^{q}}
{1-q}, \;\;\;\;\;\;\;\;\;\;\;\;\;\;\; \sum\limits_{i=1} ^{W}
p_{i}= 1,
\end{equation}
where $k$ is the Boltzmann constant and $q\in\mathbf{R}$ is the
real parameter accepting values $0<q <\infty$. In the limit $q\to
1$, we come to the well-known expression for the
Boltzmann-Gibbs-Shannon entropy, $S_{\mathrm{inf}}^{(BGS)} =
-k\sum_{i=1}^{W} p_{i}\ln p_{i}$. The information entropy
(\ref{122}) is known as Havrda-Charvat-Dar\'{o}czy-Tsallis entropy
(see~\cite{Rag99}). However, in this paper, we shall use the short
name for it. The information entropy is considered to be a measure
of uncertainty of information concerning the statistical
distribution $\{p_{i}\}$. The main its properties can be found
in~\cite{Tsal99}.

Let us introduce for further convenience a new representation for
the Tsallis information entropy with a new parameter $\xi$
\begin{equation} \label{124}
  S_{\mathrm{inf}} =k\xi\sum\limits_{i=1} ^{W} p_{i}(1-p_{i}
  ^{1/\xi}), \;\;\;\;\;\;\;\;\;\;\;\;  \xi =\frac{1} {q-1},
\end{equation}
The parameter $\xi$ takes the values $-\infty\leq\xi\leq-1$ for
$0<q\leq 1$ and $0<\xi\leq\infty$ for $1\leq q<\infty$. In
particular, in the limiting case for the value of the parameter
$q=1$, we have $\xi =\pm\infty$.

Now, the Tsallis statistical entropy in the classical and quantum
mechanics can be defined as follows
\begin{equation} \label{126}
S(t)=k\xi\int\varrho(x,p,t)[1-\varrho^{1/\xi}(x,p,t)]d\Gamma,
\;\;\;\;\;\;\;\;\;\;\; S(t)=k\xi\mathrm{Tr}
\{\varrho(t)[1-\varrho^{1/\xi}(t)]\},
\end{equation}
where $\varrho(x,p,t)$ is the phase distribution function and
$\varrho(t)$ is the statistical operator. It is easy to show that
the Tsallis statistical entropy is not additive for the fixed
value of $\xi$ and it is constant along the phase trajectories of
the dynamic system. As the total time derivative from the phase
distribution function is equal to zero, $d\varrho/dt=0$, valid
from the Liouville equation and the Liouville theorem, the total
time derivative from the classical entropy immediately yields the
equality
\begin{equation} \label{132}
  \frac{dS(t)} {dt} =
   k\xi\int\frac{d\varrho(x,p,t)} {dt}
      [1-(1 +\frac{1} {\xi})\varrho^{1/\xi}(x,p,t)]d\Gamma = 0.
\end{equation}
For the quantum ensembles, the Tsallis statistical entropy does
not depend on time. Note that for the Gibbs statistical entropy
this problem is inherent as well~\cite{Zub}.

\section{Microcanonical Ensemble}
In this section, the microcanonical distribution function and the
statistical operator will be expressed through the variables of
state of the isolated system $(E,V,z,N)$. Let us consider the
equilibrium statistical ensemble of the closed energetically
isolated systems of $N$ particles at the constant volume $V$ and
the thermodynamic coordinate $z$. It is supposed that all systems
have identical energy $E$ within $\Delta E\ll E$.

To begin with, we turn to instances of the classical case. The
Tsallis equilibrium statistical entropy~(\ref{126}) represents a
function of the parameter $\xi$ and a functional of the
equilibrium phase distribution function $\varrho_{eq}(x,p)$:
\begin{equation} \label{201}
S(\xi,\{\varrho_{eq} \})= k\xi\int\limits_{D}
\varrho_{eq}(x,p)(1-\varrho_{eq} ^{1/\xi}(x,p))d\Gamma_{N},
\end{equation}
where $d\Gamma_{N} =dxdp$ is an infinitesimal element of phase
space. Let the phase distribution function $\varrho_{eq}(x,p)$ be
distinct from zero only in the region of phase space $D$, which is
defined by inequalities $E\leq H(x,p)\leq E +\Delta E$ and be
normalized to unity:
\begin{equation} \label{202}
\int\limits_{D} \varrho_{eq}(x,p) d\Gamma_{N} =1.
\end{equation}
The phase distribution function depends on the first additive
integrals of motion of the system. In particular, it is a function
of the Hamiltonian, $\varrho_{eq}(x,p) = \varrho_{eq}(H(x,p))$.
Moreover, the Hamilton function $H(x,p)$ has the parametrical
dependence upon the number of particles $N$, volume $V$ of the
system and $z$.

For an isolated system, in the state of thermal equilibrium the
thermodynamic entropy $S_{\mathrm{T}}(E,V,z,N)$ has its maximal
value. Hence, the fundamental equation of thermodynamics
(\ref{115}) for the quasiequilibrium processes is implemented.
Changes of the variables of state at transition from one
equilibrium state to another nearby state are equal to zero,
$dE=0, dV=0, dz=0$ and $dN=0$. Therefore, from the basic equation
of thermodynamics (\ref{115}) it follows immediately that the
thermodynamic entropy at the fixed values of $E,V,z,N$ is
constant:
\begin{equation} \label{203}
(dS_{\mathrm{T}})_{EVzN} =0.
\end{equation}
To express the phase distribution function $\varrho_{eq}(x,p)$
through the variables of state $(E,V,z,N)$, let us replace the
equilibrium thermodynamic entropy $S_{\mathrm{T}}$ of the
macroscopic system with the Tsallis statistical one (\ref{201}),
$S_{\mathrm{T}}(E,V,z,N)\Longleftrightarrow
S(\xi,\{\varrho_{eq}\})$, and substitute it in Eq.~(\ref{203}).
Taking into account Eqs.~(\ref{202}) and (\ref{203}), one finds
\begin{eqnarray} \label{205}
    dS =\frac{\partial S} {\partial\xi} \ d\xi +
    \int\limits_{D}\frac{\delta S} {\delta \varrho_{eq}} \ d\varrho_{eq} \
    d\Gamma_{N} &=& 0, \\
\int\limits_{D} d\varrho_{eq} \ d\Gamma_{N} &=& 0, \label{206}
\end{eqnarray}
where the symbol $d$ before the functions $S,\xi$ and
$\varrho_{eq}$ is the total differential in variables $(E,V,z,N)$.
One should note that the unequivocal conformity between
statistical and thermodynamic entropies is satisfied for the case
where the parameters $\xi$ and $\{\varrho_{eq}\}$ are the
functions of the variables of state $(E,V,z,N)$ of the isolated
system. Let us put
\begin{equation} \label{208}
 \xi  = \frac{1} {q-1}= z.
\end{equation}
Since $d\xi=0$ and $d\varrho_{eq} =0$, we obtain from
Eqs.~(\ref{205}) and (\ref{206}),
\begin{equation} \label{209}
  \frac{\delta S(z,\{\varrho_{eq} \})} {\delta
    \varrho_{eq}} =k\alpha,
\end{equation}
where $\alpha$ is a certain constant, and $k$ is the Boltzmann
constant, which was introduced for convenience. Substituting
Eq.~(\ref{201}) into (\ref{209}), we obtain
\begin{equation} \label{210}
\varrho_{eq} ^{1/z}(x,p;E,V,z,N)=\frac{z-\alpha}{z+1}.
\end{equation}
The parameter $\alpha$ has been eliminated by using
Eqs.~(\ref{201}) and (\ref{202}):
\begin{equation} \label{211}
   \varrho_{eq}(x,p;E,V,z,N) = \left[1-\frac{S} {kz} \right]^{z} .
\end{equation}
Equations (\ref{211}) and (\ref{202}) together give
\begin{equation} \label{212}
\left[1-\frac{S} {kz} \right]^{-z} = \int\limits_{D} d\Gamma_{N} =
   \int\Delta(H(x,p)-E) d\Gamma_{N} \equiv W(E,V,N),
\end{equation}
where $\Delta(\varepsilon)$ is the function distinct from zero
only in the interval $0\leq\varepsilon\leq\Delta E$, where it is
equal to unit. The statistical weight $W(E,V,N)$ is meant as a
dimensionless phase volume, i.e., the number of dynamic states
inside a layer $\Delta E$. Based on this, we get the equipartition
probability from Eq.~(\ref{211}) as a function of the
thermodynamic ensemble variables, energy $E$,  volume $V$, number
of particle $N$, and parameter $z$~\cite{Zub}:
\begin{equation} \label{213}
 \varrho_{eq}(x,p;E,V,z,N)=W^{-1}(E,V,N)\Delta(H(x,p)-E).
\end{equation}
Thus, using Eq.~(\ref{212}), we can write the entropy as
(cf.~\cite{Tsal88,Gross})
\begin{equation} \label{214}
    S(E,V,z,N)=k z [1-W^{-1/z}(E,V,N)] =
    kz [1-e^{-S_{\mathrm{G}}(E,V,N)/kz}],
\end{equation}
where $S_{\mathrm{G}}$ is the Gibbs entropy~\cite{Gibbs,Zub} for
the microcanonical ensemble $(E,V,N)$:
\begin{equation} \label{215}
S_{\mathrm{G}}(E,V,N) = k \ln W(E,V,N).
\end{equation}

The quantum microcanonical ensemble and the corresponding
equilibrium distribution function are in some respects analogous
to the familiar classical ones. Let the probability distribution
for quantum states of the system be different from zero only in
the layer $E\leq E_{i} \leq E +\Delta E$ and be normalized to
unity:
\begin{equation} \label{216}
    \sum\limits_{i} w_{i} =1, \; \; \; \; \; \; \; \; \;
    E\leq E_{i} \leq E +\Delta E.
\end{equation}
The Tsallis equilibrium statistical entropy is a function of the
parameter $\xi$ and probabilities $\{w_{i}\}$:
\begin{equation} \label{217}
S(\xi,\{w_{i} \}) =k \xi \sum\limits_{i} w_{i}(1-w_{i} ^{1/\xi}).
\end{equation}
The repeated use of the above procedure will lead us to the
formula
\begin{equation} \label{223}
\left[1-\frac{S} {kz} \right]^{-z} = \sum\limits_{i} \Delta (E_{i} -E)
\equiv W(E,V,N).
\end{equation}
The statistical weight $W(E,V,N)$ is equal to the number of
quantum states in the layer $\Delta E$. The quantum microcanonical
distribution  becomes
\begin{equation} \label{224}
  w_{i}(E,V,z,N)=W^{-1}(E,V,N)\Delta(E_{i}-E).
\end{equation}
The statistical operator corresponding to the microcanonical
distribution of probabilities of quantum states (\ref{224}) can be
written as~\cite{Zub}
\begin{equation} \label{225}
\varrho_{eq}(E,V,z,N)=W^{-1}(E,V,N)\Delta(H-E),
\end{equation}
where the operator function $\Delta(H-E)$ is determined in the
diagonal representation by the matrix elements $\langle k
|\Delta(H-E)|k'\rangle=\Delta(E_{k} -E)\delta_{kk'}$. The quantum
statistical entropy is calculated similarly to the classical one
(\ref{214}) with statistical weight (\ref{223}). Note that the
classical and quantum microcanonical distributions (\ref{213}) and
(\ref{224}) are extreme equilibrium ones which correspond to a
maximum of the Tsallis statistical entropy~\cite{Tsal88}. The
distribution functions (\ref{213}) and (\ref{224}) obtained by the
thermodynamic method described here are identical with ones
obtained by the Jaynes principle. The index $q$ for the Jaynes
principle is a fixed parameter and does not depend on the
variables of state of the system. In this case, the Tsallis
statistics does not satisfy the zero law of
thermodynamics~\cite{Parv1}.

\section{Thermodynamics of microcanonical ensemble}
It is well-known from the conventional statistical mechanics that
in the thermal equilibrium the Gibbs entropy of the microcanonical
ensemble is an extensive variable, and it has all peculiarities of
the thermodynamic entropy in the thermodynamic
limit~\cite{Kvasn,Zub}. Mathematically, this implies that the
Gibbs entropy $S_{\mathrm{G}}$ is a homogeneous function of
variables $E,V$ and $N$ of the first order, i.e., one has the
following property~\cite{Prigogine}:
\begin{equation} \label{226}
S_{\mathrm{G}}(\lambda E, \lambda V, \lambda N) = \lambda
    S_{\mathrm{G}}(E, V, N),
\end{equation}
where $\lambda$ is a certain constant. After substitution of
Eq.~(\ref{215}) into (\ref{226}), it is easy to check up that the
statistical weight $W$ must satisfy the following requirement:
\begin{equation} \label{227}
W(\lambda E, \lambda V, \lambda N) = W ^{\lambda}(E, V, N).
\end{equation}
Taking into account Eqs.~(\ref{214}) and (\ref{226}), one finds
the following peculiarity of the Tsallis entropy
\begin{equation} \label{228}
 S (\lambda E, \lambda V, \lambda z,
\lambda N) = \lambda S (E, V, z, N),
\end{equation}
which shows that the Tsallis entropy in the microcanonical
ensemble is a homogeneous function of variables $E,V,z,N$ of the
first order. In other words, it is extensive. It is essential to
make clear that the homogeneity property of quantities
(\ref{226})-(\ref{228}) is realized only in the thermodynamic
limit.

Differentiating Eq.~(\ref{228}) with respect to $\lambda$, and
putting $\lambda=1$, we obtain the well-known Euler theorem for
the homogeneous functions:
\begin{equation} \label{229}
  E\left(\frac{\partial S} {\partial E} \right)_{V, z, N} +
    V\left(\frac{\partial S} {\partial V} \right)_{E, z, N} +
    z\left(\frac{\partial S} {\partial z} \right)_{E, V, N} +
    N\left(\frac{\partial S} {\partial N} \right)_{E, V, z} = S.
\end{equation}
Using the thermodynamic relations following from the fundamental
equation of thermodynamics (\ref{115}) in case of the isolated
thermodynamic system $(E,V,z,N)$
\begin{equation} \label{230}
    \left(\frac{\partial S} {\partial E} \right)_{V, z, N} =
     \frac{1} {T}, \; \; \;
     \left(\frac{\partial S} {\partial
     V} \right)_{E, z, N} = \frac{p} {T}, \; \; \;
      \left(\frac{\partial S} {\partial z} \right)_{E, V, N} = \frac{X} {T},
 \; \; \; \;
     \left(\frac{\partial S} {\partial
     N} \right)_{E, V, z} =-\frac{\mu} {T},
\end{equation}
we get the Euler theorem~\cite{Prigogine}:
\begin{equation} \label{231}
 T S=E+p V +X z-\mu N.
\end{equation}
Applying the differential operator with respect to the ensemble
variables $(E,V,z,N)$ on Eq.~(\ref{231}), we obtain the
fundamental equation of thermodynamics
\begin{equation} \label{232}
  T dS = dE + p dV +X dz - \mu dN
\end{equation}
and the Gibbs-Duhem relation~\cite{Prigogine}
\begin{equation}\label{233}
    S dT = V dp +z dX-N d\mu.
\end{equation}
Equation (\ref{233}) means that the variables $T$, $\mu$, $X$ and
$p$ are not independent. The fundamental equation of
thermodynamics (\ref{232}) provides the first principle
\begin{equation} \label{234}
\delta Q = dE + p dV + Xdz - \mu dN
\end{equation}
and the second law of thermodynamics
\begin{equation} \label{235}
 dS =\frac{\delta Q} {T}.
\end{equation}
Here $\delta Q$ is a heat transfer by the system to the
environment for quasistatic transition of the system from one
equilibrium state to another nearby state.

Let us investigate the homogeneity properties of the variables
$T,p,\mu$ and $X$. Substituting Eq.~(\ref{214}) into (\ref{230}),
we obtain the following expressions for the temperature
$T$~\cite{Parv1}:
\begin{equation} \label{236}
  T (E, V, z, N) = T_{\mathrm{G}}(E,
V, N) \ W ^{1/z}(E, V, N) =
   T_{\mathrm{G}}(E, V, N) \ e ^{S_{\mathrm{G}}(E, V, N)/k z}
\end{equation}
and the variable $X$
\begin{equation} \label{237}
    X (E, V, z, N) = k T_{\mathrm{G}}(E, V, N)
    [e ^{S_{\mathrm{G}}(E, V, N)/k z} -
    1-S_{\mathrm{G}}(E, V, N)/k z].
\end{equation}
The pressure and the chemical potential of the system are
equivalent with the pressure $p_{\mathrm{G}}$ and the chemical
potential $\mu_{\mathrm{G}}$ of the Gibbs statistics,
respectively, $p(E,V,z,N)=p_{\mathrm{G}}(E,V,N)$ and
$\mu(E,V,z,N)=\mu_{\mathrm{G}}(E,V,N)$. These equations were
derived by using the thermodynamic relations for the temperature
$T_{\mathrm{G}}$, the pressure $p_{\mathrm{G}}$ and the chemical
potential $\mu_{\mathrm{G}}$ of the Gibbs statistics, and taking
into account Eq.~(\ref{215}):
\begin{equation} \label{238}
\frac{1} {T_{\mathrm{G}} } = \left(\frac{\partial S_{\mathrm{G}} } {\partial
 E} \right)_{V, N}, \; \; \; \; \; \; \;
 \frac{p_{\mathrm{G}} } {T_{\mathrm{G}} } = \left(\frac{\partial S_{\mathrm{G}} } {\partial
 V} \right)_{E, N}, \; \; \; \; \; \; \;
-\frac{\mu_{\mathrm{G}} } {T_{\mathrm{G}} } = \left(\frac{\partial
S_{\mathrm{G}} } {\partial N} \right)_{E, V}.
\end{equation}
The Gibbs quantities
$T_{\mathrm{G}},p_{\mathrm{G}},\mu_{\mathrm{G}}$ are the
homogeneous functions of the variables of state $(E,V,N)$ of the
zero order. This can be proved by using Eqs.~(\ref{238}) and
(\ref{226}). Then, a combination of Eqs.~(\ref{236}) and
(\ref{226}) allows us to write the relation for the temperature
$T$:
\begin{equation} \label{241}
 T (\lambda E, \lambda V, \lambda z,
\lambda N) = T (E, V, z, N).
\end{equation}
Similarly to Eq.~(\ref{241}), the relations for the pressure
$p(\lambda E,\lambda V,\lambda z,\lambda N)$, the chemical
potential $\mu(\lambda E,\lambda V,\lambda z,\lambda N)$, and the
variable $X(\lambda E,\lambda V,\lambda z,\lambda N)$ are
fulfilled. Thus, the temperature $T$, the pressure $p$, the
chemical potential $\mu$, and quantity $X$ are the homogeneous
functions of the variables $E,V,z,N$ of the zero order. So they
are intensive variables~\cite{Prigogine}.

Let us prove in more detail the thermodynamic principle of
additivity~\cite{Kvasn}. For instance, we assume that
$\lambda=1/N$ and introduce the following specific variables:
\begin{equation} \label{242}
\varepsilon =\frac{E} {N}, \; \; \; \; \; \; \; \; \; \; \; \; \;
\; v = \frac{V} {N}, \; \; \; \; \; \; \; \; \; \; \; \; \; \; \;
\; \;
 \tilde{z} = \frac{z} {N} = \frac{1} {(q-1) N} .
\end{equation}
Thus, Eqs.~(\ref{228}) and (\ref{241}) for the entropy and the
temperature of the system, by using (\ref{242}) with respect to
$\lambda=1/N$, can be rewritten as
\begin{equation} \label{243}
s (\varepsilon, v, \tilde{z})=\frac{1} {N} \ S (E, V, z, N)
\end{equation}
and
\begin{equation} \label{244}
 T (\varepsilon, v, \tilde{z})=T (E, V, z, N),
\end{equation}
where $s(\varepsilon,v,\tilde{z})$ is the specific entropy,
$s=S/N$, which depends only on the intensive variables
$\varepsilon, v$ and $\tilde{z}$. For the pressure $p$, the
chemical potential $\mu$, and $X$, we have equations similar to
that for the temperature (\ref{244}). So, comparing
Eqs.~(\ref{243}) and (\ref{244}) with the thermodynamic equations
(\ref{106}) and (\ref{107}), we conclude that the entropy $S$ is
an extensive variable, as it is proportional to the number of
particles $N$ multiplied by an intensive variable $s$, but the
temperature $T$, the pressure $p$, the chemical potential $\mu$,
and $X$ are intensive variables.

Let us divide the system into two parts ($1$ and $2$) and require
that the total number of particles of the system should be equal
to the sum of the number of particles of each subsystem separately
and the specific quantities (\ref{242}) should be equal among
themselves
\begin{equation} \label{245}
N_{1+2}=N_{1}+N_{2}, \;\;\;\;\;\;\;
 \varepsilon_{1+2} = \varepsilon_{1} =
\varepsilon_{2},
    \; \; \; \; \; \; \;
    v_{1+2} =v_{1} =v_{2},  \; \; \; \; \; \;
   \tilde{z}_{1+2} = \tilde{z}_{1} = \tilde{z}_{2}.
\end{equation}
Then, the variables $E,V$ and $z$ are extensive. Taking into
account Eq.~(\ref{245}), one finds
\begin{equation} \label{247}
 s_{1+2}(\varepsilon_{1+2},v_{1+2},\tilde{z}_{1+2})=
 s_{1}(\varepsilon_{1},v_{1},\tilde{z}_{1})=
 s_{2}(\varepsilon_{2},v_{2},\tilde{z}_{2}).
\end{equation}
Multiplying it by the first equation from (\ref{245}) and using
(\ref{243}), we get
\begin{equation} \label{248}
 S_{1+2}(E_{1+2}, V_{1+2}, z_{1+2}, N_{1+2}) =
     S_{1}(E_{1}, V_{1}, z_{1}, N_{1}) +
      S_{2}(E_{2}, V_{2}, z_{2}, N_{2}).
\end{equation}
Thus, in the microcanonical ensemble the Tsallis entropy is an
extensive variable. Furthermore, Eqs.~(\ref{244}) and (\ref{245})
allow us to write
\begin{equation} \label{249}
   T_{1+2}(E_{1+2}, V_{1+2}, z_{1+2}, N_{1+2}) =
     T_{1}(E_{1}, V_{1}, z_{1}, N_{1}) =
      T_{2}(E_{2}, V_{2}, z_{2}, N_{2}).
\end{equation}
So, in the thermodynamic limit, the zero law of thermodynamics and
the thermodynamic principle of additivity (see. (\ref{106}) and
(\ref{107})) for the Tsallis statistics in the microcanonical
ensemble are valid. Here, the thermodynamic limit denotes the
limiting statistical procedure $N\rightarrow\infty$ at
$\varepsilon =\mathrm{const}, v =\mathrm{const}$, and
$\tilde{z}=\mathrm{const}$ with keeping the main asymptotics on
$N$. This may be explicitly seen by making an expansion of the
functions of the state in powers of the small parameter $1/N$
($N\gg 1$$, |z|\gg 1$, i.e. $q\ne 1$) with the large finite values
of the variables $E,V,z$. For the extensive functions the term
proportional to $N$ is held and for the intensive variables only
the term proportional to $N^0$ is kept (cf. Eqs.~(\ref{106}) and
(\ref{107})). Note that the correct thermodynamic limit,
$(q-1)N=\mathrm{const}$, for the Tsallis statistics has already
been discussed in Botet et al.~\cite{Botet1,Botet}. In
Abe~\cite{Abe1}, the thermodynamic limit for the Tsallis
statistics is wrong because the limits $N\to \infty$ and
$|z|\to\infty$ are not coordinated among themselves. This
procedure destroys the connection between the variables $N$ and
$z$ in the functions of the state of the system (see Section II).
In the case of the Boltzmann-Gibbs limit we make an expansion of
the functions of the state in powers of the small parameter
$1/\tilde{z}$ ($|\tilde{z}|\gg 1$, $N\gg 1$, i.e. $q\to 1$) and
hold only the zero term of the power expansion.

It is important to note that some authors~\cite{Abe0,Vives} to
make the connection with the equilibrium thermodynamics interpret
the equations similar to our Eqs.~(\ref{214}) and (\ref{236}) as
the extensive representation of the Tsallis statistics in the
terms of the physical temperature. But, as was shown
in~\cite{Parv1}, these equations are only the transformation
formulas from the Tsallis statistics to the new extensive one.

\section{The perfect gas}
The thermodynamic principle of additivity can thoroughly be
investigated in the framework of a classical nonrelativistic ideal
gas. In the microcanonical ensemble $(E,V,z,N)$, the statistical
weight (\ref{212}) of the perfect gas of $N$ identical nucleons is
given by~\cite{Das}
\begin{equation} \label{250}
 W (E, V, N) =  \frac{V^{N}}{N!}\int\frac{d^{3}p_{1}\ldots
 d^{3}p_{N}}{(2\pi\hbar)^{3N}}\
 \delta\left(\sum\limits_{i=1}^{N}\frac{\vec{p}_{i}^{2}}{2m}-E\right) =
  \frac{V^{N}}{N!}
 \left(\frac{m} {2\pi \hbar ^{2}} \right)^{\frac{3} {2}
N}  \frac{E ^{\frac{3} {2} N-1}} {\Gamma (\frac{3} {2} N)},
\end{equation}
where $m$ is the nucleon mass. In the thermodynamic limit
($N\gg1$, $\varepsilon=E/N =\mathrm{const}$,
$v=V/N=\mathrm{const}$) from Eq.~(\ref{250}), it follows
immediately that~\cite{Kvasn}
\begin{equation} \label{251}
W ^{1/N}(E,V, N) = v\left(\frac{m\varepsilon e ^{5/3}} {3\pi\hbar
^{2}}\right)^{3/2}
 \equiv w (\varepsilon, v).
\end{equation}
So Eq.~(\ref{251}) proves relation (\ref{227}) for the statistical
weight $W$ with $\lambda=1/N$. Then, the Tsallis entropy
(\ref{214}) is reduced to
\begin{equation} \label{252}
 S (E, V, z, N) = N s (\varepsilon, v, \tilde{z}),
    \; \; \; \; \; \; \; \; \; \;
 s (\varepsilon, v, \tilde{z}) = k \tilde{z} \left [1-w ^{-1/\tilde{z}}
   \right].
\end{equation}
Comparing Eq.~(\ref{252}) with (\ref{106}), we conclude that the
Tsallis entropy is an extensive variable. Note that in the limit
$\tilde{z}\rightarrow\pm\infty$, we obtain the formula for the
Gibbs specific entropy:
\begin{equation} \label{253}
\left. s (\varepsilon, v, \tilde{z}) \right |_{\tilde{z}
\rightarrow \pm\infty} =k \ln w \equiv
 s_{\mathrm{G}}(\varepsilon,v).
\end{equation}
Substituting (\ref{252}) into (\ref{230}), we get
\begin{equation} \label{254}
 T (E, V, z, N)  =
 \frac{2} {3} \frac{\varepsilon}{k} w^{1/\tilde{z}}
 \equiv T (\varepsilon, v, \tilde{z}).
\end{equation}
The temperature (\ref{254}) is a function of the specific
variables $\varepsilon, v$ and $\tilde{z}$. Therefore, it is an
intensive variable by virtue of Eq.~(\ref{107}). In the limit
$\tilde{z}\rightarrow\pm\infty$, we obtain the well-known formula
for the Gibbs statistics
\begin{equation} \label{255}
\left. T (\varepsilon, v, \tilde{z}) \right |_{\tilde{z}
\rightarrow \pm\infty} = \frac{2} {3} \frac{\varepsilon} {k} =T _
{\mathrm{G}}(\varepsilon, v).
\end{equation}
In a similar way, the pressure $p$, the chemical potential $\mu$,
and $X$ become
\begin{eqnarray} \label{256}
 p (E, V, z, N) &=&
 \frac{2} {3} \frac{\varepsilon} {v} \equiv p(\varepsilon, v, \tilde{z}), \\
\label{257}    \mu (E, V, z, N) &=&  \frac{2} {3} \varepsilon\left
[\frac{5} {2} -
 \ln w  \right] \equiv \mu (\varepsilon, v,
\tilde{z}), \\
\label{258}  X (E, V, z, N) &=& -\frac{2} {3} \varepsilon \left [1
+  \frac{1} {\tilde{z}} \ln w -
 w ^{1/\tilde{z}}  \right] \equiv X (\varepsilon, v, \tilde{z}).
\end{eqnarray}
Note that the pressure $p$ and the chemical potential $\mu$ for
the classical ideal gas in the microcanonical ensemble do not
depend on the parameter $\tilde{z}$, and they are equal to
respective quantities of the Gibbs statistics. Then,
Eqs.~(\ref{252}), (\ref{254}), and (\ref{256})-(\ref{258}) yield
the Euler theorem (\ref{231}) in terms of the specific variables
\begin{equation} \label{259}
   T s =\varepsilon +p v-\mu +X\tilde{z} .
\end{equation}
In the limit $\tilde{z} \rightarrow \pm\infty$, the pressure $p$
and the chemical potential $\mu$ remain unchanged but the variable
$X=0$. So, by the example of the classical ideal gas the principle
of additivity for the Tsallis statistics is proved. The Euler
theorem (\ref{231}) or (\ref{259}) shows that in the Tsallis
statistics the quantities $z=1/(q-1)$ and $X$ should be the
variable of the state and the associated "force", respectively.

At this point one important quantity must be noted, the heat
capacity $C_{V}=1/(\partial T/\partial E)_{V,z,N}$. In the
framework of the ideal gas of $N$ nucleons it can be written as
\begin{equation}\label{260}
    C_{V}=\frac{3}{2}k N w^{-1/\tilde{z}}
    \left( 1+\frac{3}{2}\frac{1}{\tilde{z}}\right)^{-1}.
\end{equation}
Figure 1  shows the specific heat (left) and the temperature
(center) vs the parameter $\tilde{z}$. The calculations are done
for the system of nucleons at the specific energy $\varepsilon=50$
MeV and the specific volume $v=3/\rho_{0}$, where $\rho_{0}=0.168$
fm$^{-3}$. It is of great interest that both the heat capacity and
the temperature sharply change their shape in the region of small
values of $\tilde{z}$ and considerably defer from their Gibbs
limit, which in the figure is indicated by arrows. In the region
of $-3/2 <\tilde{z}< 0$, the heat capacity is negative. It is
remarkable that such a behaviour has really been caused by the
decrease of the temperature with $\varepsilon$. This dependence
can be seen even better in right panel of Fig. 1 which shows the
temperature vs the specific energy of the system for different
values of the parameter $\tilde{z}$. The negative heat capacity in
the microcanonical ensemble is closely related to the phase
transition of the first order where the entropy is a convex
function~\cite{Chomaz,Gross1}. The Fig. 1 clearly shows that the
variable $\tilde{z}$ is the order parameter and the system is
physically unstable in the region $-3/2 <\tilde{z}< 0$ at the
fixed values of the variables $(v,\tilde{z})$, where the entropy
is a convex function of $\varepsilon$. For example, entropy $s
\sim\varepsilon^{3/2}$ at $\tilde{z}=-1$. Note that this critical
feature of the system is not related with negative values of the
parameter $q$~\cite{Tsal98} as we take into account the condition
$1+1/N\tilde{z}>0$ ($N\gg 1$). The crossing point of all curves
$(\varepsilon_{0},v_{0})$ in right panel of Fig. 1 is the point
where the Tsallis and Gibbs entropies vanish, $S=0$ and $S_{G}=0$
or $w(\varepsilon_{0},v_{0})=1$. Thus, the following values of the
energy $\varepsilon > \varepsilon_0$ and the volume $v > v_0$ of
the system have the physical sense. Note that the temperature $T$
do not depend on energy $\varepsilon$ at $\tilde{z}=-3/2$ and it
is equal with temperature $T_{0}=T(\varepsilon_{0},v_{0})$.

It should be remarked here that the microcanonical ensemble of the
Tsallis statistics in the thermodynamic limit is equivalent with
the canonical one. If we introduce the variable $z$ in the
formulas for the perfect gas in the canonical ensemble of Abe et
al.~\cite{Abe0,Abe2}, then in the thermodynamic limit we recover
the above functions of the state of the microcanonical ensemble.
For instance, in the thermodynamic limit, Eq.~(\ref{254}) can be
obtained from the energy $U_{q}(T,V,N)$ given in~\cite{Abe0,Abe2}.

\begin{figure}
  \includegraphics[width=17cm]{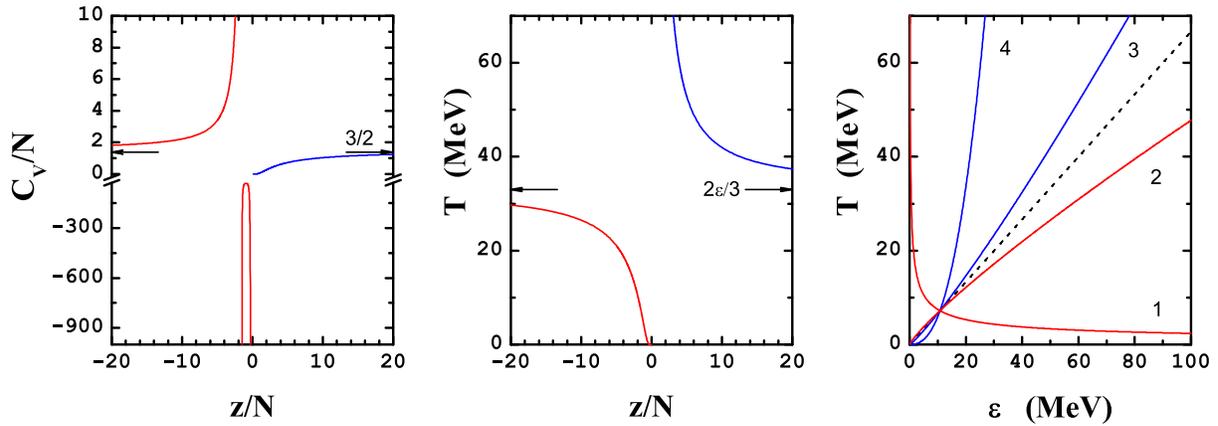}\\
  \caption{The dependence of the heat capacity (left) and the temperature (center) on
  the specific $\tilde{z}=z/N$ for the classical ideal gas of $N$ nucleons at the values of the specific energy
  $\varepsilon=50$ MeV and the specific volume $v=3/\rho_{0}$. The temperature as a function of the specific
  energy (right) for the different values of $\tilde{z}=-1$, $-10$, $10$ and $1$
  (the curves $1,2,3$ and $4$, respectively) at $v=3/\rho_{0}$. The dashed line
  corresponds to the Gibbs statistics. }\label{fg1}
\end{figure}

\section{Conclusions}
In this paper, we have explored the microscopic foundation of the
generalized equilibrium statistical mechanics based on the Tsallis
statistical entropy. The viewpoint utilized here considers that
the microcanonical ensemble is most convenient to analyze the
fundamental questions of the statistical mechanics. We summarize
our main principles.

Here, the Gibbs idea of the statistical ensembles defined within
the framework of the quantum and classical mechanics was used. In
this approach, the equilibrium phase distribution function and the
statistical operator do not depend on time, and they are functions
of the additive first integrals of motion of the system by virtue
of performance of Liouville and von Neumann equations.
Additionally, these main quantities are functions of the
macroscopic variables of state of the system. To derive the
distribution functions, in contrast with the Jaynes principle, the
new thermodynamic method based on the fundamental equation of
thermodynamics and statistical definition of the functions of the
state of the system was given.

In this paper, we have made the following claim. The index $\xi$
of the Tsallis entropy should be an extensive variable of the
state of the system. As a result of this assumption, we obtain
that in the microcanonical ensemble the Tsallis entropy represents
the homogeneous function of the variables $E,V,z,N$ of the first
order. The temperature of the system is an intensive variable,
and, consequently, the zero law of thermodynamics is satisfied.
Other functions of state of the system are either extensive or
intensive. Thus, in the thermodynamic limit,
$\tilde{z}=1/(q-1)N=\mathrm{const}$, in the Tsallis statistics the
thermodynamic principle of additivity is carried out. Note that
the Tsallis information entropy is nonextensive because the
parameter $\xi$ is a certain intensive constant. Also it is
necessary to note that the Tsallis statistical entropy as well as
the Gibbs one has an essential lack. Both the entropies do not
depend on time while the thermodynamic entropy grows up to achieve
its maximal value in the state of thermal equilibrium. The
extensive property of the Tsallis entropy in the microcanonical
ensemble yields the Euler theorem which permits one to find the
fundamental equation of thermodynamics and the Gibbs-Duhem
relation. Thus, the first and the second principles of
thermodynamics are fulfilled. Note that in the limit,
$\tilde{z}\rightarrow\pm\infty$, all expressions of the Tsallis
statistics take the form of the conventional Gibbs statistical
mechanics. So the Tsallis statistical mechanics in the
microcanonical ensemble satisfies all postulates of the
equilibrium thermodynamics.

Finally, the classical nonrelativistic ideal gas of $N$ identical
nucleons in the microcanonical ensemble was considered to
illustrate the principles which were elucidated in the general
theory. It has been shown that in the thermodynamic limit the
statistical weight, the entropy, the temperature, and other
quantities are the homogeneous functions of the first and zero
order of the variables of state, respectively. Note that for ideal
gas the Euler theorem was accomplished and in the limit,
$\tilde{z}\rightarrow\pm\infty$, all expressions resembled ones of
the Gibbs statistics.

{\bf Acknowledgments:} This work has been supported by the
Moldavian-US Bilateral Grants Program (CRDF project MP2-3045). We
acknowledge valuable remarks and fruitful discussions with
T.S.~Bir\'{o}, R.~Botet, K.K.~Gudima, M.~P{\l}oszajczak, and
V.D.~Toneev.


\end{document}